\documentclass[prb,amsmath,twocolumn,amssymb,floatfix,showpacs,superscriptaddress,longbibliography]{revtex4-1}

\usepackage{epstopdf}
\usepackage{graphicx}
\usepackage{amsmath}
\usepackage{amssymb}
\usepackage{mathbbol}
\usepackage{braket}
\usepackage{svg}
\usepackage{xcolor}
\usepackage{epsfig}

\usepackage[utf8]{inputenc}
\usepackage[english]{babel}

 \usepackage[colorlinks=true,linktoc=page,linkcolor=magenta,citecolor=blue,urlcolor=purple]{hyperref}
 
\usepackage{ulem}

\usepackage[utf8]{inputenc}

\makeatother

\begin{document}

\graphicspath{{Figs/new/}}
	
%%%%%%%%%%%%%%%%%%%%%%%%%%%%%%%%%%%%%%%%%%%%%%%%%%%%%%%%%%%%%%%%%%%%%%%%%%%%%%%%%%%%%%%%%%%%%%%%
%% Title
%%%%%%%%%%%%%%%%%%%%%%%%%%%%%%%%%%%%%%%%%%%%%%%%%%%%%%%%%%%%%%%%%%%%%%%%%%%%%%%%%%%%%%%%%%%%%%%%
	
\title{Electronic confinement of surface states in a topological insulator nanowire} 
	
%%%%%%%%%%%%%%%%%%%%%%%%%%%%%%%%%%%%%%%%%%%%%%%%%%%%%%%%%%%%%%%%%%%%%%%%%%%%%%%%%%%%%%%%%%%%%%%%
%% Authors
%%%%%%%%%%%%%%%%%%%%%%%%%%%%%%%%%%%%%%%%%%%%%%%%%%%%%%%%%%%%%%%%%%%%%%%%%%%%%%%%%%%%%%%%%%%%%%%%
	
\author{Ruchi Saxena}
\affiliation{Advanced Technology Institute and Department of Physics, University of Surrey, Guildford, GU2 7XH, United Kingdom}
	
\author{Eytan Grosfeld}
\affiliation{Department of Physics, Ben-Gurion University of the Negev, Beer-Sheva 8410501, Israel}

\author{Sebastian E de Graaf}
\affiliation{National Physical Laboratory, Hampton Road, TW11 0LW, Teddington, United Kingdom}

\author{Tobias Lindstrom}
\affiliation{National Physical Laboratory, Hampton Road, TW11 0LW, Teddington, United Kingdom}

\author{Floriana Lombardi}
\affiliation{Quantum Device Physics, Chalmers University of Technology, Sweden}
	
\author{Oindrila Deb}
\affiliation{Advanced Technology Institute and Department of Physics, University of Surrey, Guildford, GU2 7XH, United Kingdom}	
	
\author{Eran Ginossar}
\affiliation{Advanced Technology Institute and Department of Physics, University of Surrey, Guildford, GU2 7XH, United Kingdom}
	
\date{\today}
	
%%%%%%%%%%%%%%%%%%%%%%%%%%%%%%%%%%%%%%%%%%%%%%%%%%%%%%%%%%%%%%%%%%%%%%%%%%%%%%%%%%%%%%%%%%%%%%%%
%% Abstract
%%%%%%%%%%%%%%%%%%%%%%%%%%%%%%%%%%%%%%%%%%%%%%%%%%%%%%%%%%%%%%%%%%%%%%%%%%%%%%%%%%%%%%%%%%%%%%%%
	
\begin{abstract}
\noindent
	We analyze the confinement of electronic surface states in a model of a topological insulator nanowire. Spin-momentum locking in the surface states reduces unwanted backscattering in the presence of non-magnetic disorder and is known to counteract localization for certain values of magnetic flux threading the wire. We show that intentional backscattering can be induced for a range of conditions in the presence of a nanowire constriction. We propose a geometry for a nanowire that involves two constrictions and show that these regions form effective barriers that allow for the formation of a quantum dot. We analyze the zero-temperature non-interacting electronic transport through the device using the Landauer-B\"{u}ttiker approach and show how externally applied magnetic flux parallel to the nanowire and electrostatic gates can be used to control the spectrum of the quantum dot and the electronic transport through the surface states of the model device.
\end{abstract}
	
\maketitle
	
%%%%%%%%%%%%%%%%%%%%%%%%%%%%%%%%%%%%%%%%%%%%%%%%%%%%%%%%%%%%%%%%%%%%%%%%%%%%%%%%%%%%%%%%%%%%%%%%
%% Section: Introduction
%%%%%%%%%%%%%%%%%%%%%%%%%%%%%%%%%%%%%%%%%%%%%%%%%%%%%%%%%%%%%%%%%%%%%%%%%%%%%%%%%%%%%%%%%%%%%%%%
%\textit{Introduction.}--- 
\section{Introduction}
\label{introduction}

\noindent
Topological insulators (TI) are bulk insulators that exhibit surface states with unique electronic properties. Three-dimensional (3D) TI nanowires (TINWs) are promising candidates for studying the electronic surface transport \cite{tinr1,tinr2,tinr3,Berry2}; they admit two-dimensional (2D) helical surface states that prohibit backscattering and reduce localisation in the absence of magnetic disorder while reducing any residual bulk transport due to their geometry \cite{sym1,sym2,sym3,sym4,sym5}.  While this protection from unintended scattering harbours promise, designing mesoscopic scale structures that can exploit the topological protection  does require control of the electronic transport, including methods to control transmission and to generate electronic confinement (e.g., quantum dots, quantum point contacts). Successful control of charge transport would pave the way for various applications of TIs ranging from improved spintronic devices, through more accurate charge pumps for quantum metrology to quantum computing \cite{qd_appl1,qd_appl2,qd_appl3,schmitt2020}.

There have been various theoretical proposals for inducing a gap in the surface spectrum of the TIs since this would allow to achieve the electronic confinement and control needed for practical devices. These include exploiting the exchange coupling induced by a proximitised magnetic insulator \cite{gapped1}, by means of surface pairing through a proximitised superconductor \cite{gapped2}, or realising a tunnel coupling between the top and bottom surfaces of an ultra-thin TI \cite{gapped3,gapped4,gapped5,gapped6}. The latter method has been experimentally explored in a quantum dot fabricated from a thin film of a 3D TI with a gate tunable barrier. It was found by varying the gate voltage that the charge transport ranged from ohmic to tunneling regimes \cite{gapped7}, although signatures of quantised levels localised within the dot region remained elusive. On the other hand, geometrically constructed quantum devices for charge confinement have not been fully explored in TINWs.\\

 TINWs are promising with regards to realising full electronic confinement, and progress in understanding transport was made in recent works which studied effects of disorder, ripples and magnetic fields \cite{Hamil_tinr1,Hamil_tinr2,Hamil_tinr3,Hamil_tinr4}. Recently, signatures of the sub-bands of the TI nanowire have been detected experimentally \cite{exp_CB_Ando} creating additional motivation to analyse the potential of such devices for quantum confinement along the nanowire. 

In this work we show, by analysing the conditions for quantum confinement within a defined section of a long TINW, that a quantum dot can be created which transport is uniquely based on the nature of the topological surface states. The emergence of the quantum dot in our work is strictly a quantum mechanical effect which is based on the quantum interference of the Dirac surface states. The analysis is based on an effective cylindrical model for a microstructure with defined radius variations and external electric and magnetic potentials. A key element is a reduced radius region (constriction) exhibiting an increased gap in the spectrum of surface states. We show that this region acts like a potential barrier which can backscatter Dirac electrons \cite{Hamil_tinr2}. Conditions for charge confinement can be found when backscattering in the TINW is allowed between non-time reversal symmetric states as well as for cases where backscattering is facilitated by the dynamics in the constrictions where time-reversal symmetry (TRS) is broken.  The radius variation is essential when the TINW is flux-biased to the gapless state of half-flux for which the surface states are topologically protected from backscattering by scalar potentials. This is considered to be the point of optimal performance due to the resilience against disorder in the leads. Based on this, we propose a geometrical construction that can lead to resonant tunnelling \cite{Tsu_1973} and form discrete states within a confined central region, see Fig.~\ref{nrnrn_model}. We show that the geometry exhibits clear evidence of quantum dot formation that is manifested in the electronic transport as sub-gap resonances. %through the device. 
% \od{Moreover, we find that away from half-flux, sub-gap resonances can form for a uniform nanowire in the presence of a gate- induced electrostatic double barrier potential and no constrictions.}

The remainder of this paper is organised as follows. In section \ref{proposed geometry}, we present the geometrical construction of a TINW for a quantum dot and discuss the scattering between incoming and reflected states that can lead to finite reflection. In section \ref{Surface Hamiltonian and scattering analysis}, we describe the Hamiltonian and the scattering matrix for the model device to compute the conductance employing the Landauer-B\"{u}ttiker approach. We also investigate the effect of the curved interface on the motion of the surface states by incorporating the spin connection which is essential to analyze the motion of Dirac particle on a curved space time \cite{Hamil_tinr1,Berry3,spincon1,spincon3}. We discuss the numerical results in section \ref{numerical results} and conclude our findings in section \ref{conclusion}.

% We employ the scattering matrix approach to compute the scattering amplitudes and find the conductance through the model device using the Landauer-B\"{u}ttiker formalism. We investigate the effect of the curved interface on the motion of the Dirac fermions by incorporating the spin connection \cite{Hamil_tinr1,Berry3,spincon1,spincon2,spincon3}. We show that the quantised energy levels in the dot are tunable by the application of the external flux in parallel to the wire and electrostatic potentials. 

%%%%%%%%%%%%%%%%%%%%%%%%%%%%%%%%%%%%%%%%%%%%%%%%%%%%%%%%%%%%%%%%%%%%%%%%%%%%%%%%%%%%%%%%%%%%%%%%
%% Section: Proposed geometry
%%%%%%%%%%%%%%%%%%%%%%%%%%%%%%%%%%%%%%%%%%%%%%%%%%%%%%%%%%%%%%%%%%%%%%%%%%%%%%%%%%%%%%%%%%%%%%%%
%\textit{Proposed geometry.}--- 
\section{Proposed geometry}
\label{proposed geometry}

\begin{figure}[!]
	\begin{center}
		%\includesvg[width=5in,height=2.5in]{nrnrn_subtrate1.svg}
\hspace*{-0.5cm}\includegraphics[width=3.5in,height=3.3in]{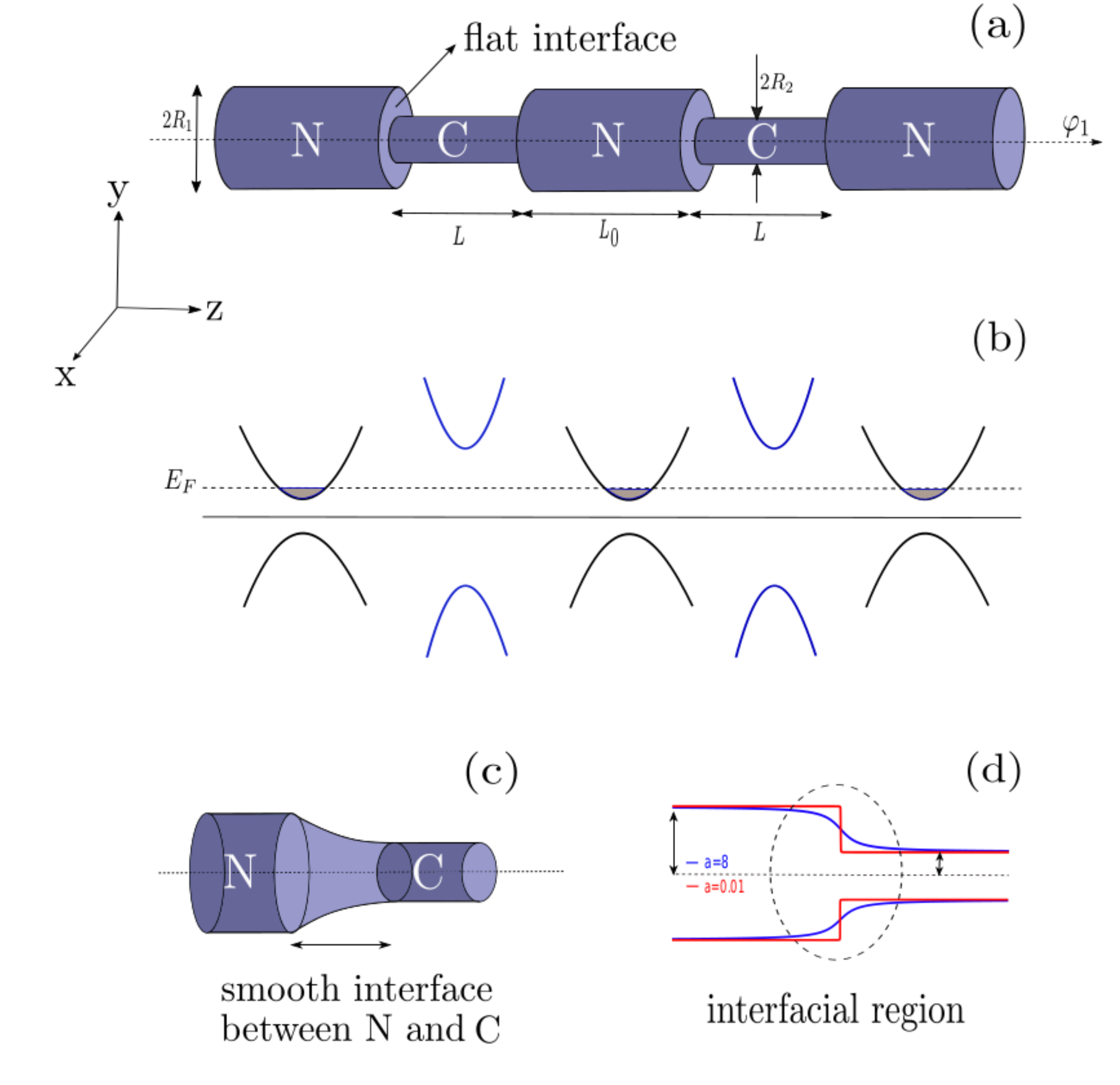}
	\end{center}
	\caption{\hspace*{-0.1cm} Geometrical model of the quantum dot based on a 3D TINW:  (a) The geometry used for the theoretical study of the transport properties. N and C denote regions of radius $R_1$ and the reduced radius regions of radius $R_2$ formed by etching the wire respectively. The interfaces between N and C regions are drawn flat for simplicity. (b) The bandgap profile (as explained in the main text) experienced by an incident electron. (c) A smooth interface connecting the N and C regions of the TINW. (d) The radius dependence of the TINW as function of the coordinate $z$ along the wire with the parameter $a=8$ corresponding to (c) and $a=0.01$ generating a step-like interface.}
	\label{nrnrn_model}
\end{figure}

\noindent
We start by presenting the proposed geometry for the TINW quantum dot (Fig.~\ref{nrnrn_model}a). We consider a cylindrical nanowire of radius $R_1$, which is etched at two regions to a radius $R_2$ ($R_2<R_1$).  We label this device NCNCN where ``N'' refers to the region with radius $R_1$ and ``C'' denotes the region with the reduced radius $R_2$. Both regions are assumed to have full rotational-symmetry around the wire axis. The wave functions in each region satisfy an anti-periodic boundary condition around the nanowire perimeter due to the curvature-induced $\pi$ Berry phase. As detailed in section \ref{Surface Hamiltonian and scattering analysis}, due to the radius dependent band gap in the TINW \cite{Yeyati,Arijit}, the band-gap profile experienced by the incoming electronic states is expected to behave as shown schematically in Fig.~\ref{nrnrn_model}b. Etching is expected to have some gradual radius profile, as depicted in Fig.~\ref{nrnrn_model}c. To incorporate its effect on the electronic transport we model the interface using a $z$-dependent radius, as demonstrated in Fig.~\ref{nrnrn_model}d, with
\begin{align}
	R(z) = R_1 + (R_2-R_1)F(z),
	\label{radialF}
\end{align}
where $F(z)=\frac{1}{2} \left[1+\frac{2}{\pi}\tan^{-1}\left(\frac{z}{a}\right)\right]$ is the smooth Heaviside theta function with the parameter $a$ tuning the interface from step-like (small $a$) to smooth (large $a$).

% Experimentally, it is challenging to etch the nanowire uniformally around the circumference and make an annulus like interface (flat) between the N and C regions as shown in Fig.~\ref{nrnrn_model}a. However, there has been experimental studies on nanostructures of 3D TIs which are etched to ultrathin films \cite{gapped7}. These studies suggest an alternative to the cylindrical nanowire geometry which can be straightforwardly fabricated.

% In time-reversal symmetric TINW, degenerate surface states wrap around the surface of the cylindrical wire forming transverse modes, and two quantum numbers label each band: (i) the momentum along the wire axis $k$; and, (ii), the total angular momentum $l$, which labels the transverse modes. Here $l$ is a half-integer due to the presence of a $\pi$ spin Berry phase acquired by the $2\pi$ rotation of the spin around the cylindrical surface \cite{Berry1,tinr4,Berry3}. 

Controlled reflection of the electrons from the interfaces is essential to realising charge confinement. The motion of the particle belonging to a specific sub-band can be thought of as one-dimensional (only dispersing along the wire axis) with the gap proportional to the angular momentum $l$. The incoming and reflected states of same $l$ are not Kramers pairs and hence, are prone to backscattering due to disorder \cite{Yeyati,Hamil_tinr2,disorder}. The unusual finite reflection here is strikingly different than the zero reflection (unit transmission) situation where backscattering is prohibited due to the spin-momentum locked gapless and orthogonal surface states. The scattering between non-orthogonal surface states in TINW, characterised by $k$ of opposite sign and same $l$,  can be thought to be a result of the finite angular momentum. Similar scattering is paradigmatic to the suppression of Klein tunnelling in graphene \cite{graphene_th1, graphene_expth1} and in TIs \cite{Diptiman,Oindrila} when a massless Dirac fermion is incident at an oblique angle to the potential barrier. A finite transverse component of the momentum arising from the oblique incidence allows backscattering of the incident Dirac fermions \cite{graphene_exp1,graphene_exp2,graphene_exp3,graphene_exp4, Oindrila} due to the non-orthogonal incoming and reflected states.
%##############  Fig 1  #####################
%###########################################
%%%%%%%%%%%%%%%%%%%%%%%%%%%%%%%%%%%%%%%%%%%%%%%%%%%%%%%%%%%%%%%%%%%%%%%%%%%%%%%%%%%%%%%%%%%%%%%%
%% Section: Surface hamiltonian
%%%%%%%%%%%%%%%%%%%%%%%%%%%%%%%%%%%%%%%%%%%%%%%%%%%%%%%%%%%%%%%%%%%%%%%%%%%%%%%%%%%%%%%%%%%%%%%%
%\emph{Surface hamiltonian.}---
\section{Surface Hamiltonian and scattering analysis}
\label{Surface Hamiltonian and scattering analysis}

\noindent
The Hamiltonian for the surface states of a non-uniform 3D TINW can be derived using a field theoretic approach \cite{Hamil_tinr1,Hamil_tinr2,Hamil_tinr3,Hamil_tinr4}. We provide the derivation of the Hamiltonian in the Appendix for notational convenience. To include the coaxial  magnetic field we use the symmetric gauge to write the vector potential $A_\phi=-Br/2=-\Phi/2\pi r$, where $\phi$ denotes the azimuthal direction, r is the radial coordinate, and $\Phi$ is the magnetic flux along the nanowire axis. With the minimal coupling of the magnetic field to the transverse motion of the Dirac surface states the Hamiltonian becomes
\begin{equation}
	\begin{split}
		H &= \hbar v_F \Bigg[ \frac{1}{\sqrt{1+{R'(z)}^2}} \left\{i  \partial_z + \frac{i  R'(z)}{2R(z)}\right\} \sigma_y  \\
		&\hspace{3cm}+\frac{1}{R(z)}\left(-i\partial_\phi - \varphi \right) \sigma_z \Bigg],
	\end{split}
	\label{H3}
\end{equation}
where $R(z)$ is the radial function which is defined to model the interface between N and C regions and $\varphi=\Phi/\Phi_0$ is the dimensionless magnetic flux. The term proportional to $R'$ is the spin-connection contribution to the Dirac fermions moving along the curved surface \cite{Hamil_tinr1,Hamil_tinr2,Hamil_tinr3,Hamil_tinr4}. This Hamiltonian is particularly useful for finding the appropriate boundary condition at the curved interface between the normal and reduced radius regions. We note that we only consider the orbital effect of the applied magnetic field because at low magnetic field (e.g. $\varphi=0.5$), the Zeeman energy is negligible compared to the energy spacing of the sub-bands \cite{Zeeman_Mason}. \\

For the cylindrical surfaces in the N and C regions, away from the interface, $R'=0$. For this case, exploiting the rotational symmetry of the cylinder and translational invariance along the wire axis we can write the solution of the time-independent Schr\"{o}dinger equation for this Hamiltonian as 
\begin{equation}
	\Psi_{k,l}(z,\phi)= e^{ik z} \, e^{i l \phi} \, \psi_{k,l},
	\label{ansatz}
\end{equation}
here $\psi$ is a two component spinor and $l$ is a half-integer angular momentum due to the presence of a $\pi$ spin Berry phase acquired by the $2\pi$ rotation of the spin around the cylindrical surface \cite{Berry1,Berry2,Berry3}. Using the solution given in Eq.\ref{ansatz} the Hamiltonian in N and C regions can be written as
\begin{equation}
	\begin{split}
		H_{1,2} &= \hbar v_F \Bigg[-k \sigma_y+\left(\frac{l-\varphi_{1,2}}{R_{1,2}}\right) \sigma_z \Bigg].
	\end{split}
	\label{H4}
\end{equation}
The energy of the surface states can be found by diagonalizing the above Hamiltonian, 
\begin{equation}
    \nonumber E_{1,2}=\pm \hbar v_F \sqrt{k^2_{1,2}+(l-\varphi_{1,2})^2/R^2_{1,2}},
\end{equation}
where $1,2$ refers to N and C region respectively. Due to the reduced cross section in the C region, the flux penetrating into this region is given by $\varphi_2=\varphi_1 R_2^2/R_1^2$. \\
\vspace*{0.2cm}
%##############  Fig 2  #####################
\begin{figure}[!ht] 
	\centering  
	\includegraphics[width=2.4in,height=1.8in]{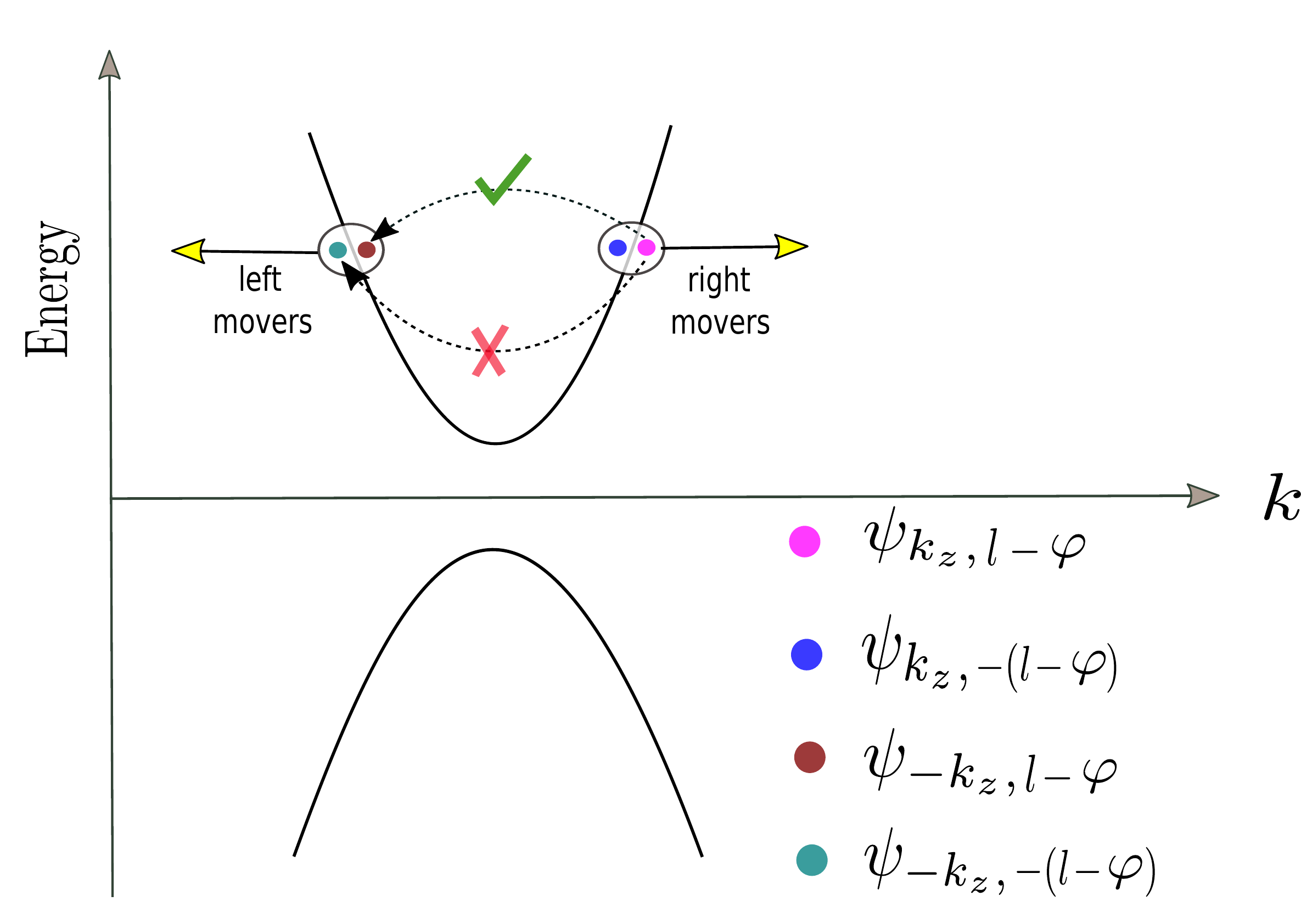}  
	\caption{ The allowed and forbidden scattering between orthogonal ($\psi_{k,l-\varphi}$ (pink colour) and $\psi_{-k,-(l-\varphi)}$ (green colour)) and non-orthogonal states ($\psi_{k,l-\varphi}$ (pink colour) and $\psi_{-k,l-\varphi}$ (brown colour)) is shown schematically.}  
	\label{scattering}
\end{figure}

We now explain the scattering processes in Fig.~\ref{scattering} where we schematically show the allowed and forbidden scattering between the right and left moving spinors. Note that the magnetic flux threaded along the wire axis breaks TRS and all the surface states become non-degenerate. However, TRS can be restored at integer and half-integer values of magnetic flux and all bands, except the $l=\varphi$ band become degenerate \cite{Berry3, TRsym_tinr1, Hamil_tinr3, Yeyati}. Now, if we fix the Fermi energy within the bulk gap such that it crosses the lowest degenerate sub-band (which happens at integer values of magnetic flux), there always exists two right moving spinors $\psi_{k_{1},\pm (l-\varphi)}$ and two left moving spinors $\psi_{-k_{1},\pm (l-\varphi)}$ as demonstrated in Fig.~\ref{scattering}. As discussed in section \ref{introduction}, although the backscattering is prohibited between the time-reversal partner $\psi_{k_{1},l-\varphi}$ and $\psi_{-k_{1},-(l-\varphi)}$, it is allowed between two other available non-orthogonal states $\psi_{k_{1},l-\varphi}$ and $\psi_{-k_{1},l-\varphi}$. As a result, the non-magnetic disorder can backscatter the Dirac surface states. 

\begin{figure}[!ht]
	\begin{center}
		\hspace*{-0.3cm}\includegraphics[width=2.3in,height=2.4in]{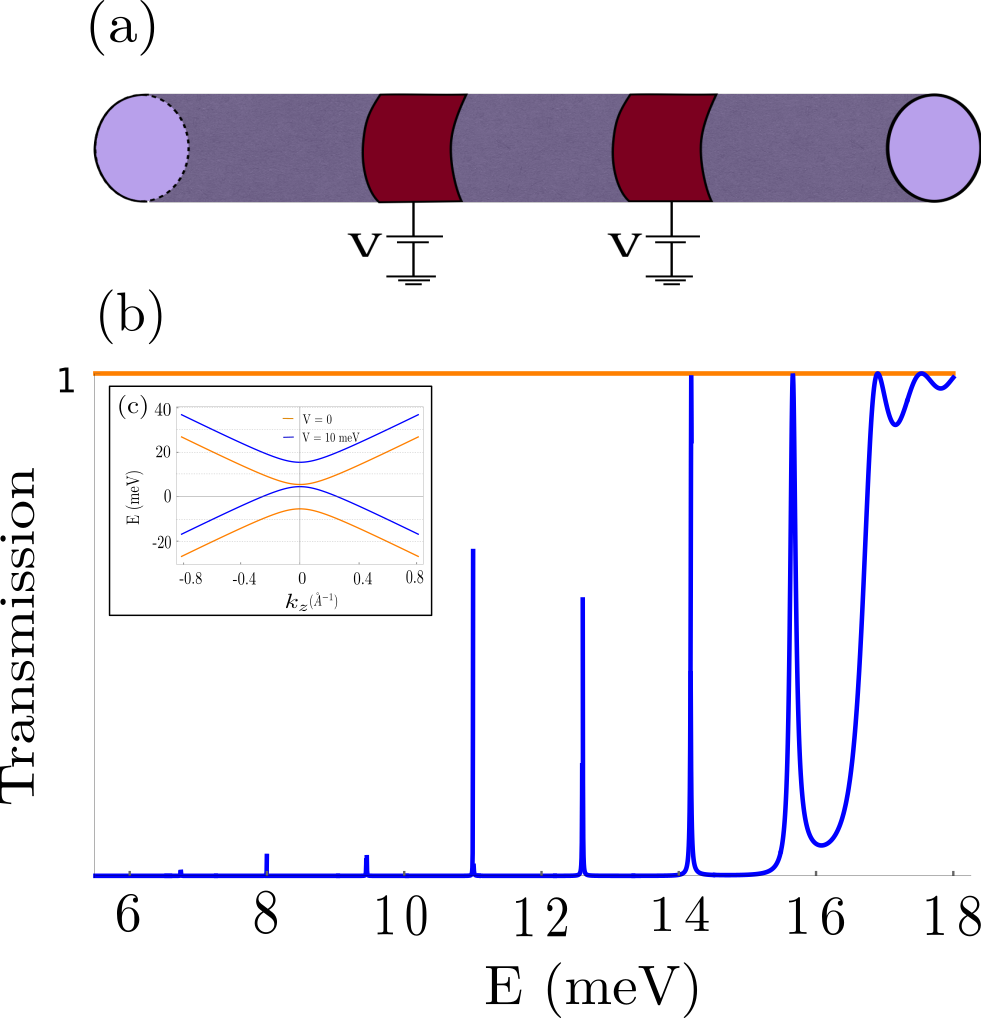}\\
	\end{center}
	\caption{\hspace*{0cm}(a) The uniform TINW in presence of two electrostatic potentials $V$. (b) If the energy of the incoming state is fixed between the conduction band for V=10 meV and V=0 sub-gap transmission resonances are found within the region between the potential barrier. The inset shows the dispersion of the $l=1/2$ band in a uniform nanowire of radius 30 nm for zero flux in the absence (orange colour) and presence (blue colour) of a potential barrier.}
	\label{disorder}
\end{figure}

First, we analyse a uniform nanowire for the case of zero magnetic flux. We consider the scattering of surface states belonging to the $l=1/2$ band in presence of a double barrier electrostatic potential $V$ induced by gates in a TINW with uniform radius, see Fig.~\ref{disorder}a. Notice that due to a non-zero potential, there is a uniform upward shift in the energy spectrum as shown in Fig.~\ref{disorder}b which creates a finite energy difference between the conduction bands belonging to the section with potential V and the section without the potential. By fixing the energy of the incoming surface state within this energy window, we compute the transmission through the TINW for zero magnetic flux and find sub-gap (the gap between the conduction band for V=10 meV and V=0 in Fig.~\ref{disorder}b) resonances in the transmission probability function of the surface states through the TINW as shown in Fig.\ref{disorder}c, which is a typical signature of bound state formation within the middle region. The formation of bound states indicate the fact that surface states in a TINW can backscatter from a scalar disorder (non-magnetic disorder).
\\

%##############  Fig 3 #####################

For a half-integer value of flux along the nanowire, there is always a corresponding value of $l$ for which the surface state is non-degenerate and
gapless \cite{Hamil_tinr3,Yeyati}. In this particular state, an electrostatic barrier as shown in Fig. \ref{disorder}a, can not backscatter the Dirac particles. Thus, in order to generate confinement we need an additional mechanism to ensure a finite gap in the constrictions in the TINW system. Hence, we propose to have reduced radius regions in the TINW so that the flux penetrating these regions is not a half-integer as opposed to N region. Consequentially, the surface states at these constricted regions become gapped due to the breaking of TRS. To summarize, constricted region ensures a larger band gap between the surface states as compared to N regions for both with and without magnetic flux.\\
\hspace*{0.5cm}The external flux control on the electronic transport through the clean NCNCN geometry is a unique feature of the TINW quantum dot where each C region has radius $R_2<R_1$. The finite reflection due to breaking of TRS in C regions resembles the backscattering in quantum spin Hall edge in the presence of local magnetic disorder that breaks the TRS \cite{Diptiman}.
Furthermore, the external gate tunability of the band gap in the constricted region can provide another probe to control the electronic transport through the model device. In section \ref{numerical results}, we discuss both external magnetic flux and gate dependent transport in detail.\\
% for both integer and half-integer values of flux.
% To summarize, in both the cases of uniform TINW and constricted TINW, bound states will arise though the mechanism of constructing the potential barrier like regions is different.\od{what do you mean by mechanism? not very clear }\\
\hspace*{0.5cm} Now, we present the scattering matrix approach to find the scattering coefficients for the non-interacting electrons. Let us consider the interface at $z=0$ between an N and a C region in the presence of a finite flux along the nanowire axis. We compute the transfer matrix that connects the wave function from the left side of the interface  ($\psi (z<0)$) to the right side of it  ($\psi (z>0)$). This can be done by solving the Schr\"{o}dinger equation $H \psi=E\psi$, where H is given by Eq.(\ref{H4}). By integrating the Schr\"{o}dinger equation across the interface we arrive at the following wave function matching condition (for the curved interface as shown in Fig.\ref{nrnrn_model}c)
\begin{equation}
	\psi(z>0)=T_z \, \psi (z<0),
	\label{bc}
\end{equation}
where the $2\times2$ transfer matrix $T_z$ is given by 
\begin{equation}
	\begin{split}
		T_z &= \mathcal{P}_z \, \text{Exp}\Bigg[ \int_{z_0}^{z_1}  \Big\{\frac{-R'}{2R} -\frac{\sqrt{1+R'^2}}{R}\Big(l-\varphi\Big)\sigma_x \\
		&\hspace*{3.5cm}- \frac{iE \sqrt{1+R'^2}}{\hbar v} \sigma_y\Big\} dz \Bigg],
	\end{split}
	\label{tmatrix}
\end{equation}
here $\mathcal{P}_z$ is the path ordered product of the exponential factor along the wire.\\

% %############## Fig 4 ################
% \begin{figure}[!ht]
% 	\begin{center}
% 		\includegraphics[width=2.2in,height=2.4in]{nr2.png}
% 		%\includegraphics[width=3.4in,height=1.8in]{interface1.png}
% 	\end{center}
% 	\caption{\hspace*{-0.1cm} (a) Curved interface of the etched nanowire and (b) Theoretical modelling of (a) using smooth Heaviside theta function as explained in Eq.(\ref{radialF}). Here, a (in units of nm) is a parameter that models curved (blue) and flat (red) interface for its large and small values respectively.}
% 	\label{nr_interface}
% \end{figure}

\hspace*{-0.4cm}Next, considering the elastic scattering within the device we can write the wave functions in each region of the model device shown in Fig.\ref{nrnrn_model}a as
\begin{eqnarray}
	\psi_{1,3,5} &=& \frac{a_{1,3,5}}{\sqrt{2}} 
	\left({\begin{array}{c} 
			\frac{-\chi_{l} + E}{i k_{1}} \\ 1
	\end{array}}\right)e^{ik_{1} z}
	+
	\frac{b_{1,3,5}}{\sqrt{2}} 
	\left({\begin{array}{c} 
			\frac{-\chi_{l} + E}{-i k_{1}} \\ 1
	\end{array}}\right) e^{-ik_{1} z} \nonumber\\
	\psi_{2,4} &=& a_{2,4}
	\left({\begin{array}{c} 
			\frac{-\chi_2+ E}{i k_2} \\ 1
	\end{array}}\right) e^{ik_2 z} 
	+ 
	b_{2,4}
	\left({\begin{array}{c} 
			\frac{-\chi_2 + E}{-i k_2} \\ 1
	\end{array}}\right) e^{-ik_2 z},
	\label{wfs}
\end{eqnarray}
here $\psi_{1,3,5}$ and $\psi_{2,4}$ are the wave functions in the normal N and reduced radius regions C respectively and $k_{1,2}=\sqrt{E^2-{\chi_{1,2}}^2}$, $\chi_{1,2}=(l-\varphi_{1,2})/R_{1,2}$.  $a_{1,2,3,4,5}$ and $b_{1,2,3,4,5}$ are the scattering coefficients. Here 1,3,5 refers to N regions while 2,4 indicates C regions. We use the wave function matching condition given in Eq.(\ref{bc}) to find the total transfer matrix M that connects the incoming states in the leftmost N region to the outgoing states in the right most N region and we get 
\begin{eqnarray}
	M = \alpha_1^{-1}\, T_z\, \beta_1\,\alpha_2^{-1}\, T_z^\dag \, \beta_2 \, \alpha_3^{-1} \, T_z \, \beta_3\, \alpha_4^{-1} \, T_z^\dag\, \beta_4,
	\label{masterM}
\end{eqnarray}
where $\alpha_{1,2,3,4}$ and $\beta_{1,2,3,4}$ are given as
\begin{eqnarray}
	\alpha_i &=& \left({\begin{array}{cc} 
			\frac{-\chi_i + E}{i k_i} e^{ik_i z_i} &  \frac{-\chi_i + 
				E}{- i k_i} e^{-ik_i z_i}\\
			e^{ik_i z_i}  & e^{-ik_i z_1} 
	\end{array}}\right) \nonumber\\ %alpha1
	\beta_i &=& \left({\begin{array}{cc}
			\frac{-\chi_i + E}{\sqrt{2} i k_i} e^{ik_i z_i} &  \frac{-\chi_i + 
				E}{-\sqrt{2} i k_i} e^{-ik_i z_i}\\ 
			e^{ik_i z_i}  &  e^{-ik_i z_i} 
	\end{array}}\right) \nonumber \\ %beta1
	\alpha_3 &=& \alpha_1 \quad \quad (\text{at}\,  z_1=z_3) \nonumber \\ %alpha3
	\alpha_4 &=& \alpha_2 \quad \quad (\text{at}\,  z_2=z_4) \nonumber \\ %alpha4
	\beta_3 &=& \beta_1 \quad \quad (\text{at}\,  z_1=z_3) \nonumber \\ %beta3
	\beta_4 &=& \beta_2 \quad \quad (\text{at} \, z_2=z_4), %beta4
	\label{scatteringM}
\end{eqnarray}
Here $i=1,2,3,4,5$ as before. The inverse of the matrix element $M_{11}$ gives the transmission probability which we use in the next section to discuss the characteristics of the quantum dot device proposed in this work.  \\

%%%%%%%%%%%%%%%%%%%%%%%%%%%%%%%%%%%%%%%%%%%%%%%%%%%%%%%%%%%%%%%%%%%%%%%%%%%%%%%%%%%%%%%%%%%%%%%%
%% Section: Numerical results
%%%%%%%%%%%%%%%%%%%%%%%%%%%%%%%%%%%%%%%%%%%%%%%%%%%%%%%%%%%%%%%%%%%%%%%%%%%%%%%%%%%%%%%%%%%%%%%%
%\textit{Numerical results} %Scattering formalism
\section{Numerical results}
\label{numerical results}
\begin{figure}[!ht]
	\begin{center}
		\includegraphics[width=3.5in,height=1.6in]{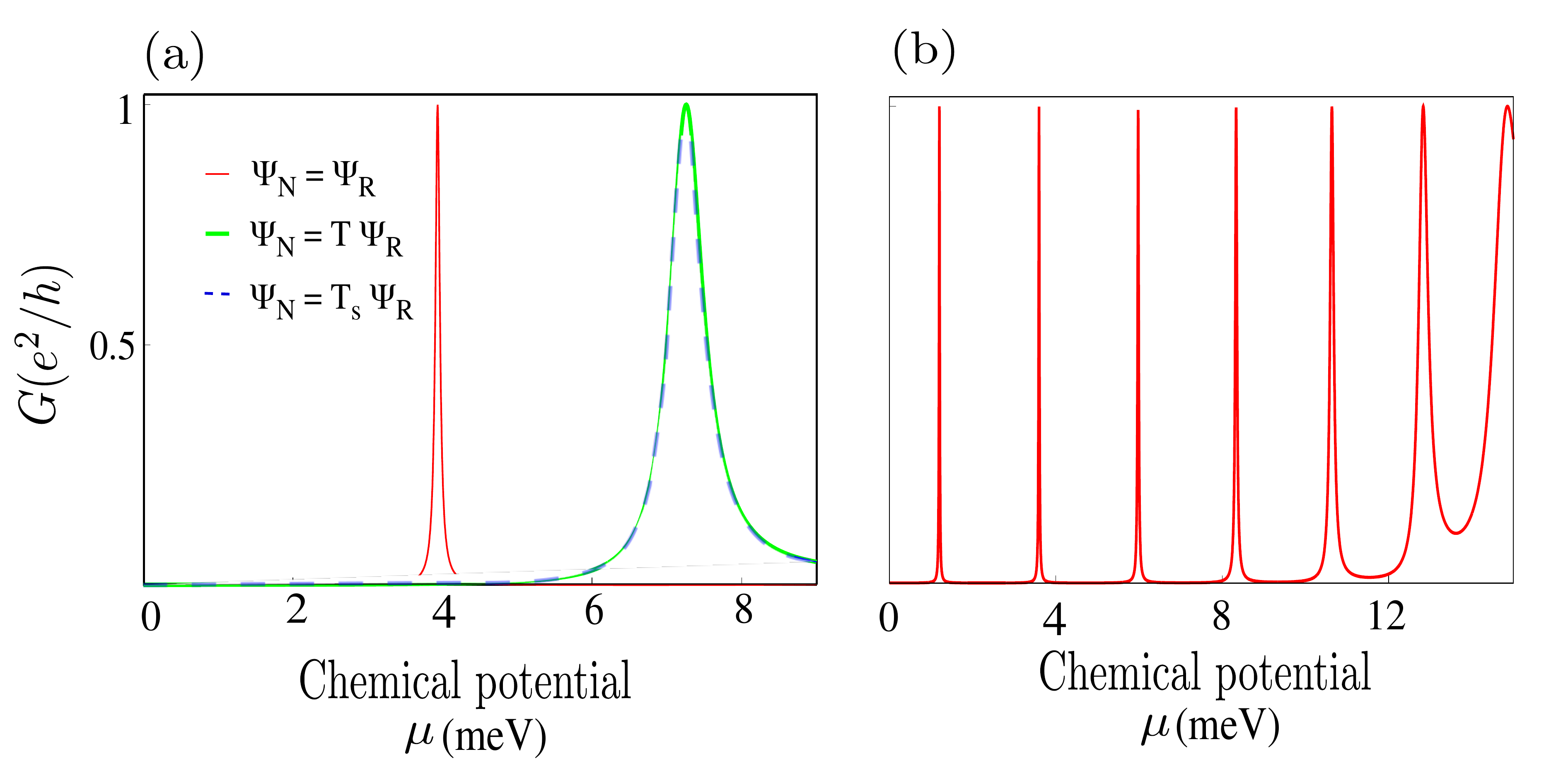}
	\end{center}
	\caption{\hspace*{-0.1cm} (a) Conductance through an etched TI nanowire with flat interfaces (as sketched in Fig.(\ref{nrnrn_model}a)) is plotted as a function of chemical potential $\mu$ for a dot length $L_0=100\, \text{nm}$. Red colour depicts the result for flat interface. The position of conductance peaks not only shifts but also broadens in case of curved interface as shown by the blue (dotted) and green (solid) curves (here, former and later colour represent the result with and without incorporating spin connection for the curved manifold). It is due to the modified boundary conditions (see, Eq.(\ref{tmatrix})). More detail about the boundary conditions in all three cases is discussed in the main text. (b) Sub-gap resonant transmission through the long quantum dot $L_0= 400$ nm in TI nanowire NCNCN geometry having flat interfaces. The numerical values of the system parameters are chosen as ${\text{R}}_1=80$ nm, ${\text{R}}_2=15$ nm, ${\text{L}}=80$ nm, $\varphi_1=0.5$\,\, \text{and}\,\, $\varphi_2=\varphi_1 {\text{R}}_2^2/{\text{R}}_1^2$.}
	\label{trans_nrnrn}
\end{figure}

\noindent
We now discuss the zero-temperature electronic transport through the nanowire structure. We consider the energy of the incoming states such that they are propagating modes in N regions (with radius $R_1$) and the radius of C regions is always smaller than $R_1$. The propagating states with total angular momentum $l=1/2$ coming from the left side of the device can have a finite reflection at the first interface between N and R. This is due to non-orthogonality of the incoming state $\psi_{k_1, (l-\varphi)}$ and the outgoing states $\psi_{-k_1, (l-\varphi)}$ which are not the time reversal partner.
Since the C regions have larger gap than in N, these constricted regions effectively behave like the potential barrier for the Dirac particles and thus can scatter the incoming particles. The multiple scattering events in the middle region of NCNCN give rise to resonance effects which manifest in the total transmission function through the device. We employ the Landauer-B\"{u}ttiker approach \cite{Supriyo} to compute the conductance through the NCNCN device at zero temperature and plot it in Fig.\ref{trans_nrnrn}a (red colour) as a function of the chemical potential $\mu$ for the flat interface which corresponds to the boundary condition with $T_z=1$. In experiments, $\mu$ is influenced by substrate, contacts etc. and can be locally tuned using electrostatic gates. The conductance peaks correspond to the sub-gap resonances in the transmission probability through the model device and exhibit clear evidence
of bound states formation inside the middle N region of the device. Notice that the N regions are immune to scalar disorder for $\varphi=l$ due to appearance of gapless surface states. More generally, when the dimensions of the middle section is small compared to the mean free path, we do not expect disorder to significantly affect the resonances.

%##############  Fig 4  #####################

\vspace{0.5cm}

%##############  Fig 5  #####################
\begin{figure}[!ht]
	\begin{center}
		\includegraphics[width=3.5in,height=1.4in]{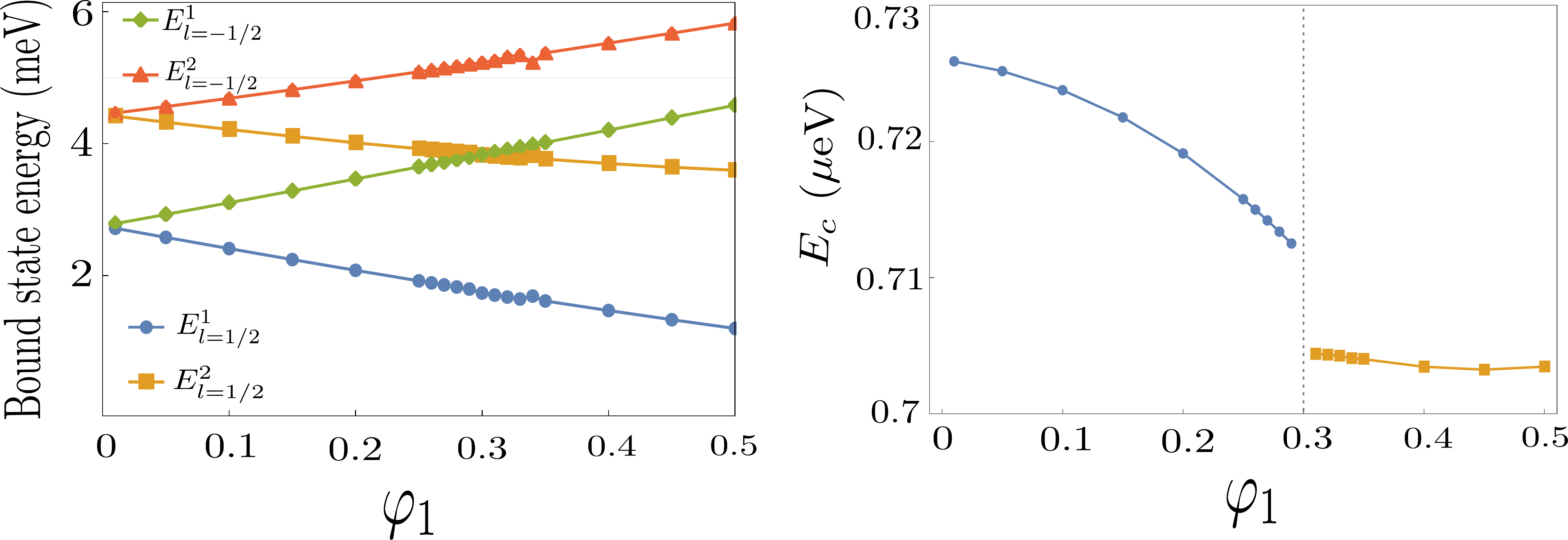}
	\end{center}
	\caption{\hspace*{-0.1cm} \textcolor{black}{(a) The lowest two bound state energies as a function of the magnetic flux $\varphi_1$ for $l=1/2$ and $l=-1/2$. (b) Quantum contribution to the Coulomb energy ($E_c^q$) as a function of the magnetic flux. The dotted line in (b) is drawn to represent the degeneracy point. Other parameters are the same as in Fig.\ref{trans_nrnrn}b.}}
	\label{cb_vsFlux}
\end{figure}

% The NCNCN model for $\phi=l=1/2$ is in contrast with the uniform nanowire in presence of two gate voltages (Fig.~\ref{disorder}) as discussed in previous section where none of the regions are protected against any accidental disorder.\\

The position of the conductance peaks in Fig.\ref{trans_nrnrn}a gives the energy of the bound states. The width of a conductance peak depends on the length $L$ of the C regions while the number of these peaks depends on the length $L_0$ of the middle N region  (the confined region) where we find the bound states. The value of the model parameters used in this work are chosen as ${\text{R}}_1=80 \,\text{nm},{\text{R}}_2=15 \,\text{nm}, \text{L}=80 \,\text{nm}, {\text{L}_0}=100\,\, \text{nm}\,\,\, \text{and}\, \varphi_1=0.5$ which are experimentally attainable \cite{tinr1,tinr2,exp_CB_Ando}. Surface sensitive experiments on $\text{Bi}_2\text{Se}_3$ estimate the value of the coherence length to be 500 nm and a typical value of the mean free path to be 100 nm \cite{exp_CB_Ando,Zeeman_Mason,tinr1}. Therefore, the electronic transport in the TINR studied in this work is expected to be ballistic and phase coherent. It is important to note that device parameters can be found for the case with one non-degenerate sub-gap resonance. In that case, only one electron can be inside the dot. However, there will be multiple quantum dot states for a larger quantum dot. We show the transmission through the device for a larger dot of length $L_0=400$ nm and same radius $R_2=15$ nm in Fig.~\ref{trans_nrnrn}b which exhibits many sub-gap resonances in the transmission function. An useful quantity in the discussion of quantum dot is the charging energy ${\text{E}}_{\text{c}}={\text{e}}^2/\text{C}_0$, where $C_0$ is the capacitance that depends on the device dimensions and the dielectric constant of the material. The value of $C_0$ can be inferred from the experiments based on  $\text{Bi}_2\text{Se}_3$ nanowires. For a $\text{Bi}_2\text{Se}_3$ nanowire on a $\text{SiO}_2/\text{Si}$ substrate $C_0=2 \times 10^{-17}$ F for a surface area $8.6 \times 10^{-14}\, \text{m}^2$ was found \cite{gapped7}, giving an effective capacitance $C_{\text{eff}}=2.3 \times 10^{-4} \, \text{F}\, \text{m}^{-2}$. We can now estimate the charging energy of the quantum dot (${\text{R}}_2=15$ nm, ${\text{L}}_0$=400nm) using \cite{Hamil_tinr1}, $\text{E}_{\text{c}}={\text{e}}^2/2\pi {\text{R}}_1 {\text{L}}_0 {\text{C}}_{\text{eff}}$ and this gives 3.6 meV. I.e. Coulomb blockade like oscillations of the conductance are expected in transport experiments on a large dot  by varying the gate voltage applied to the central NCNCN region. However, as we pointed out earlier, the origin of the discrete quantum dot states here uniquely arises from interference of the surface states, different from the appearance of quantum dot based on Coulomb blockade. \cite{Beenakker1991,McEuen1991} \\

\hspace*{0.7cm} Next, we examine the effect of a curved interface. We utilize the radial function given in Eq.(\ref{radialF}) of the section \ref{proposed geometry} to incorporate the curvature effects on the electronic transport. Here, we use the boundary condition, $\psi(z>0)= {\text{T}}_z \, \psi (z<0)$ where ${\text{T}}_z$ is the transfer matrix for $R'=0$. It is clear from Fig.\ref{trans_nrnrn}a (blue curve) that not only the position of the conductance peak gets shifted but also the peak broadens compared to the situation when we consider the flat interface. Furthermore, we analyse the motion of the Dirac particle on the curved interface incorporating the spin connection in the transfer matrix ($R'\neq0$). We plot the conductance again in Fig. \ref{trans_nrnrn}a (green curve) which coincides with the plot computed for $R'=0$ (blue colour). The findings suggest that the electronic transport is unaffected by the inclusion of the spin connection.\\

%##############  Fig 6  #####################
\begin{figure}[!ht]
	\begin{center}
		\includegraphics[width=3in,height=2.3in]{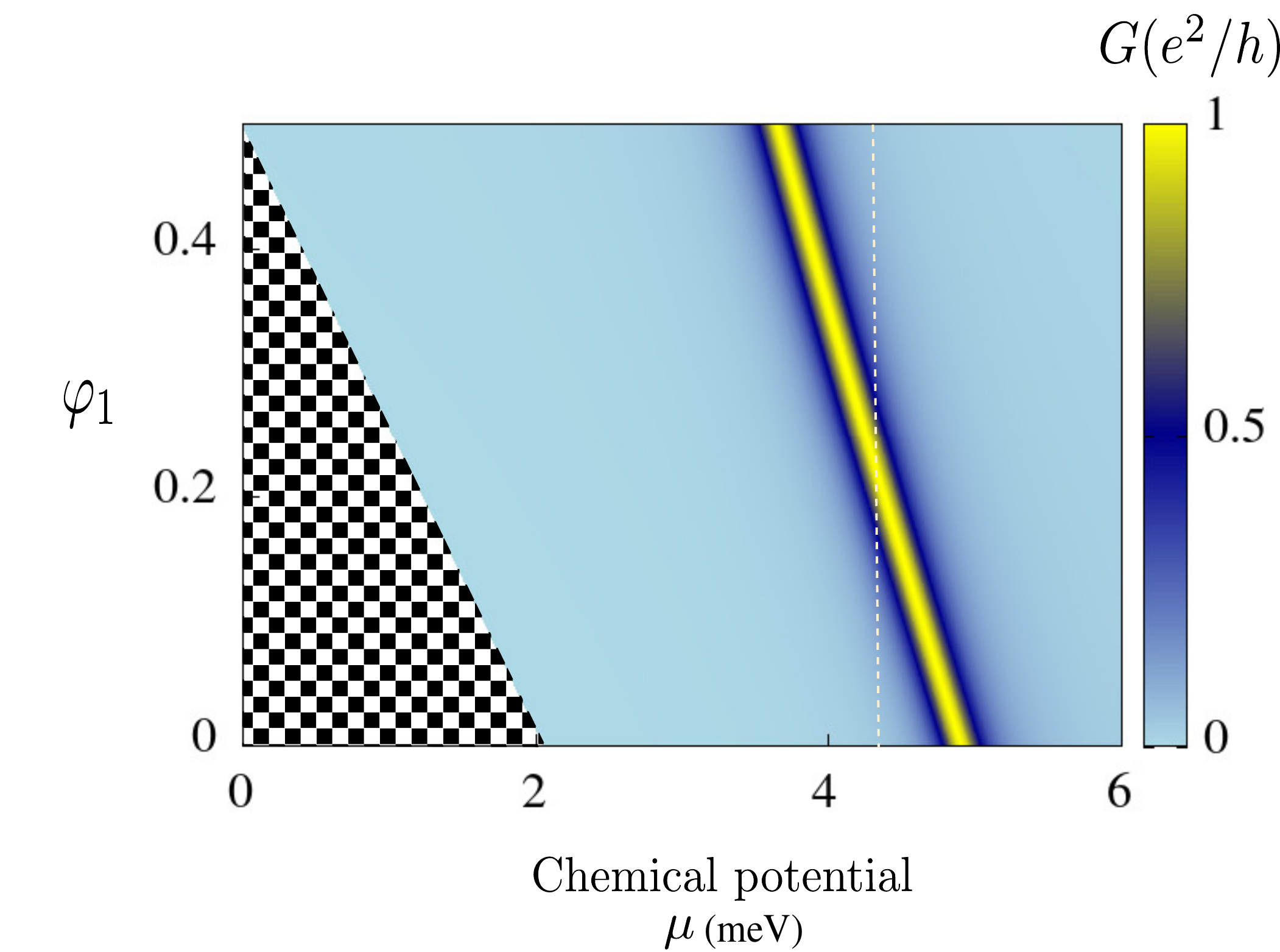}
	\end{center}
	\caption{\hspace*{-0.1cm} Conductance as a function of chemical potential $\mu$ and the flux along the wire axis $\varphi_1$. Device parameters are fixed at ${\text{R}}_1=80\, \text{nm}, {\text{R}}_2=20\, \text{nm}, \text{L}=80 \, \text{nm} \,\, \text{and} \,\, {\text{L}}_0=100 \, \text{nm}$. The checker board region is non-physical because of the evanescent modes in N region.}
	\label{tvs_phi_E}
\end{figure}

%##############  Fig 7  #####################
\begin{figure}[!ht]
	\begin{center}
		\includegraphics[width=3in,height=2.3in]{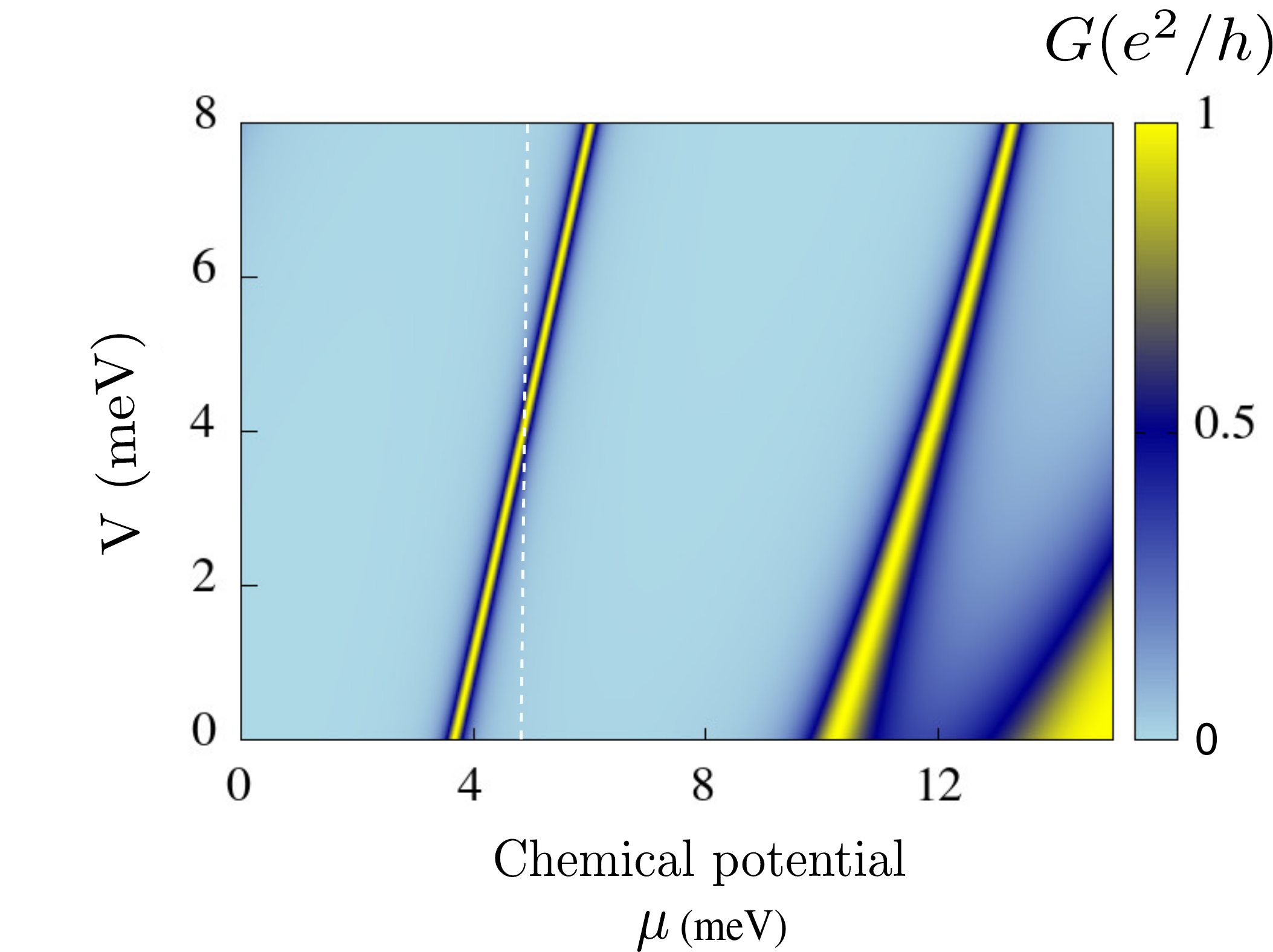}
    \end{center}
	\caption{\hspace*{-0.1cm} External gate tunability of the conductance is shown as a function of the electric potential induced on the dot by the electrostatic gates applied in the C regions and the chemical potential $\mu$. The values of other parameters are taken as in Fig.\ref{tvs_phi_E}.}
	\label{tvs_E_V} 
\end{figure}

Apart from the classical charging energy of the formed quantum dot, which arises from the long range Coulomb potential, we also estimate the quantum contribution $\text{E}_\text{c}^\text{q}$ due to the screened short-range Coulomb interaction \cite{ALEINER2002} which depends on the quantum dot states. We study its dependence on the magnetic flux. In Fig.\ref{cb_vsFlux}a, we show how the energy of the bound states varies as a function of magnetic flux $\varphi_1$. We represent the lowest two bound states corresponding to the $l=1/2$ and $l=-1/2$ bands as $E_{l=1/2}^1,\, E_{l=1/2}^2$ and $E_{l=-1/2}^1, \,E_{l=-1/2}^2$ respectively in Fig.\ref{cb_vsFlux}a. We model the short-range interaction as a contact interaction and compute the Coulomb energy between electrons which occupy the two lowest energy states. For $\varphi_1<0.3$, the two occupied states correspond to $E_{l=1/2}^1$ and $E_{l=-1/2}^1$ while $E_{l=1/2}^1$ and $E_{l=1/2}^2$ become the lowest energy states for $\varphi_1>0.3$. We plot the Coulomb energy $\text{E}_\text{c}^\text{q}$ in Fig.\ref{cb_vsFlux}b as a function of flux $\varphi_1$. We observe a smooth decay in $\text{E}_\text{c}^\text{q}$ as the magnetic flux increases up until the point of degeneracy at $\varphi_1=0.3$. As flux increases beyond 0.3 we notice slow decrease in $\text{E}_\text{c}^\text{q}$. We note that we do not estimate $\text{E}_\text{c}^\text{q}$ at one point $\varphi_1=0.3$ as the second lowest quantum dot state is degenerate, see Fig.\ref{cb_vsFlux}a.

The bandgap tunability of the surface state by an external flux enables us to investigate the electronic transport as a function of the flux threaded along the wire axis. We plot the conductance through the model device in Fig.~\ref{tvs_phi_E} for the flat interfaces between the N and C regions as a function of chemical potential $\mu$ and the flux $\varphi_1$ applied along the wire axis in the N region. Evidently, if the chemical potential is fixed (consider the dotted white line in the plot) the transport of the surface states can be externally controlled by the application of a flux along the wire.  The checker board region in Fig.~\ref{tvs_phi_E} corresponds to the presence of evanescent modes in the N region and hence is non-physical. Next, we examine the gate voltage tunability of the electronic transport in presence of half-integer flux. The two gate voltages in C regions change the band gap profile by shifting the band spectrum (with respect to the band spectrum in N regions) upward by the electrostatic potential V applied by the gate. We plot the conductance as a function of induced gate potential and the chemical potential in Fig.~\ref{tvs_E_V} for flux $\varphi_1=0.5$ and we find that the conductance can be controlled by varying the gate voltage. \\

\begin{figure}[!ht]
	\begin{center}
		\includegraphics[width=3in,height=2.3in]{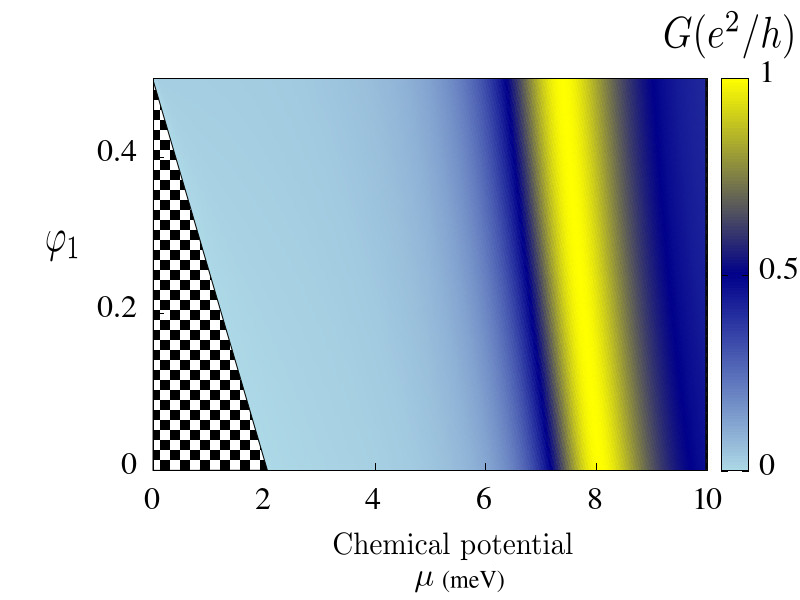}
    \end{center}
	\caption{\hspace*{-0.1cm} \textcolor{black}{The conductance through the NCNCN quantum dot model considering the smooth interface (taking a=8 nm, as shown in Fig.\ref{nrnrn_model}d) as a function of chemical potential and magnetic flux $\varphi_1$. The values of all parameters are the same as in Fig.\ref{tvs_phi_E}.}}
	\label{tvs_phi_E_smooth} 
\end{figure}

\begin{figure}[!ht]
	\begin{center}
		\includegraphics[width=3in,height=2.3in]{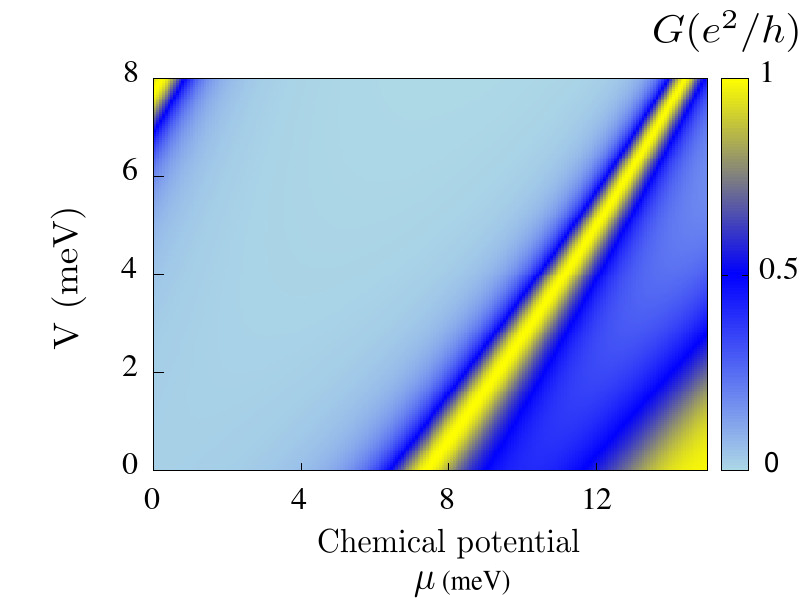}
    \end{center}
	\caption{\hspace*{-0.1cm} \textcolor{black}{ The conductance through the NCNCN quantum dot model considering the smooth interface (the same as for Fig.\ref{tvs_phi_E_smooth}) as a function of chemical potential and external gate voltage. The values of all parameters are the same as in Fig.\ref{tvs_phi_E}.}}
	\label{tvs_E_V_smooth} 
\end{figure}

\hspace*{0.5cm}So far, we study the quantum dot considering the flat interfaces between different sections of the device. However, while etching/pattering nanowire the interfaces are likely to undergo geometrical deformation. As discussed earlier in section \ref{proposed geometry}, to address the transport in this situation, we model the interface as shown in Fig.\ref{nrnrn_model}d and assume a  smoothness $a=8$ nm, a relatively smooth interface that is obtainable using modern nanofabrication techniques, yet significantly smaller than other dimensions of the NCNCN structure. We investigate the influence of such interfaces on the electronic transport as a function of external flux, gate voltages and chemical potential. First, we show the conductance as we vary flux and chemical potential in Fig.\ref{tvs_phi_E_smooth}. We observe that the resonance line broadens and gets shifted on the energy axis. The broadening of the conductance peak relates to the inverse of the life time of the quantum dot state which means that the particle would quickly escape the dot. Similar features are found when we plot conductance as a function of chemical potential and the gate voltage in Fig.\ref{tvs_E_V_smooth}. 
These results can be attributed to the strong coupling of the dot to the left and right N regions which determines the shape and position of the conductance peaks. Thus, from an experimental point of view it is important to engineer as sharp interfaces as possible.

%##############  Fig 8 #####################

%##############  Fig 9  #####################

The external control of quantum transport is a necessary prerequisite in the fabrication of field-effect transistors and other quantum devices. The external magnetic flux as well as induced gate voltage tunability makes our geometrical construction of the TI quantum dot experimentally feasible for experimental implementation. Moreover, we anticipate that the transmitted current through the quantum dot states in this device might have different spinor state for different conditions fixed by chemical potential, magnetic flux and voltage bias. We leave the thorough study of characterising the finite temperature current in presence of Coulomb interaction and disorder for degenerate and non-degenerate dot for future work.

%\newpage
%                   DISCUSSION AND CONCLUDING REMARKS
%\textit{Discussion and concluding remarks} 
\section{Conclusions}
\label{conclusion}
\noindent
In this work, we exploited the finite reflection processes for designing a model for a quantum dot in a TINW. We showed that finite reflection is possible essentially due to the transverse angular momentum modes. We demonstrated that in the presence of double barrier potential, the uniform nanowire can host bound states if the Fermi energy lies within the conduction band minimum for $V\neq0$ and $V=0$ for a fixed value of angular momentum $l$. We elucidated the necessity of the constricted region in presence of half-integer magnetic flux along the wire axis in N region (half-integer magnetic flux is important to prevent the backscattering due to accident disorder in N region). In particular, we propose a theoretical model of a quantum dot in a TINW exploiting the unique characteristics of the surface states and show that the confining potential due to the larger gap in the surface states in constricted regions localise the Dirac surface state. Finally, we explicitly show that the electronic transport through discrete quantum dot states can be modulated by external magnetic flux and gate voltage. This will facilitate some of the most desirable quantum devices for metrology (single-electron devices, charge pumps), spintronics (spin-polarized current source), quantum optics, and quantum computing. 

% This work provides analysis of how to utilise the geometry of the nanowire and the band structure of the surface states to achieve the required quantum confinement which will facilitate these device

%                   ACKNOWLEDGEMENTS

\acknowledgments{We acknowledge support from the European Union's Horizon 2020 Research and Innovation Programme under Grant Agreement No. 766714/HiTIMe.}

\vspace{1cm}
\appendix
\textit{Appendix: Hamiltonian on the curved surface}\\

\noindent
The surface of the TINW can be described using the space time metric in 2+1D ($z,\phi,t$)
\begin{equation}
	g_{\mu,\nu}=(-1,1+{{R}'}^2,R^2),
	\label{metric1}
\end{equation}
where $R'$ is the change in radius along the nanowire. This metric is related to the Minkowski metric for the flat surface by the following relation 
\begin{equation}
	\begin{split}
		g_{a,b} = g_{\mu,\nu}\, {e^\mu}_a \, {e^\nu}_b,
		\label{metric2}
	\end{split}
\end{equation}
Here Latin letters ($a,b$) denote the local coordinates and Greek letters ($\mu,\nu$) denote general coordinates. 
The Dirac equation in curved space is
\begin{equation}
	\begin{split}
		i \gamma^\mu \, D_\mu \phi &= 0,
		\label{DEq1}
	\end{split}
\end{equation}
where the covariant derivative, $D_\mu=\partial_\mu + \omega_{ab\mu}\, \Sigma^{ab}/4$.
It can be shown that only $\omega_{12\phi}$ and $\omega_{21\phi}$ are non zero \cite{Hamil_tinr3} and given by 
\begin{equation}
	\omega_{12\phi}=\omega_{21\phi}=\frac{{R}'}{\sqrt{1+{{R}'}^2}},
\end{equation}
so the covariant derivatives are given by 
\begin{equation}
	\begin{split}
		D_t & = \partial_t,  \\
		D_z & = \partial_z,  \\
		D_\phi & = \partial_\phi + \omega_{12\phi} \, \Sigma^{21}/2. 
	\end{split}
	\label{covD}
\end{equation}
Using these expressions we can write the Dirac Eq.(\ref{DEq1}) as 
\begin{equation}
	\begin{split}
		i \gamma^t \partial_t \psi + i \gamma^z \partial_z \psi + i \gamma^\phi \left[\partial_\phi +\omega_{12\phi} \Sigma^{21}/2  \right] \psi &= 0,
		\label{DEq2}    
	\end{split}
\end{equation}
where
\begin{equation}
	\begin{split}
		\Sigma^{21} &= \frac{1}{2}\left[\Gamma^1,\Gamma^2\right] \quad \text{and} \quad \gamma^\mu \equiv {e^\mu}_a \, \Gamma^a.
		\label{sigmagamma}
	\end{split}
\end{equation}
To calculate vierbein or tetrad, $(e^\mu)_a$ we use Eq.(\ref{metric2}) and find three set of equations 
\begin{equation}
	\begin{split}
		g_{00} & = g_{\mu \nu}  \, {e^\mu}_0 \, {e^\nu}_0, \\
		g_{11} & = g_{\mu \nu} \, {e^\mu}_1 \, {e^\nu}_1 ,\\
		g_{22} & = g_{\mu \nu} \, {e^\mu}_2 \, {e^\nu}_2 ,
	\end{split}
\end{equation}
which can be expended in terms of the components of $g_{\mu\nu}$ as
\begin{equation}
	\begin{split}
		-1 & = -({e^t}_0)^2 + (1+{{R}'}^2)({e^z}_0)^2 + R^2 ({e^\phi}_0)^2, \\
		-1 & = -({e^t}_1)^2 + (1+{{R}'}^2)({e^z}_1)^2 + R^2 ({e^\phi}_1)^2,\,  \\
		-1 & =-({e^t}_2)^2 + (1+{{R}'}^2)({e^z}_2)^2 + R^2 ({e^\phi}_2)^2,   
	\end{split}
\end{equation}
now taking only the following tetrad non-zero
\begin{equation}
	\begin{split}
		{e^t}_0 & = 1, \\
		{e^z}_1 & = \frac{1}{\sqrt{1+{{R}'}^2}}, \\
		{e^\phi}_2 & = \frac{1}{R}, \\
	\end{split}
\end{equation}
we find $\gamma^\mu$ using Eq.(\ref{sigmagamma})
\begin{equation}
	\begin{split}
		\gamma^t & \equiv {e^t}_a \Gamma^a = {e^t}_0 \Gamma^0 + {e^t}_1 \Gamma^1 +{e^t}_2 \Gamma^2 = \Gamma^0,\\
		\gamma^z & \equiv {e^z}_a \Gamma^a = {e^z}_0 \Gamma^0 + {e^z}_1 \Gamma^1 +{e^z}_2 \Gamma^2 = \frac{1}{\sqrt{1+{{R}'}^2}}\Gamma^1,\\
		\gamma^\phi & \equiv {e^\phi}_a \Gamma^a = {e^\phi}_0 \Gamma^0 + {e^\phi}_1 \Gamma^1 +{e^\phi}_2 \Gamma^2 = \frac{1}{R}\Gamma^2.\\        
	\end{split}
\end{equation}
We can put these results in Eq.(\ref{DEq2}) and get
\begin{equation}
	\begin{split}
		i \Gamma^0 \partial_t \psi + i \frac{1}{\sqrt{1+{{R}'}^2}} \Gamma^1 \partial_z \psi + i \frac{1}{R} \Gamma^2 \Bigg[\partial_\phi\\
		+ \frac{R'}{\sqrt{1+{{R}'}^2}} \Sigma^{21}/2\Bigg] \psi = 0.
	\end{split}
	\label{DEq3}
\end{equation}
Now, we choose the following Dirac matrices 
\begin{equation}
	\begin{split}
		\Gamma^0 & = i \sigma_x,\\
		\Gamma^1 & = i \sigma_y,\\
		\Gamma^2 & = i \sigma_z,
	\end{split}
\end{equation}
and write Eq.~\eqref{DEq3} as
\begin{equation}
	\begin{split}
		-\sigma_x \partial_t\psi + i \frac{1}{\sqrt{1+{R'}^2}} \sigma_z \partial_z \psi + \sigma_y \frac{i}{R} \Bigg[\partial_\phi \\ 
		+ \frac{R'}{\sqrt{1+{R'}^2}} \Sigma^{21}/2\Bigg]\psi = 0,
	\end{split}    
\end{equation}
where $$\Sigma^{21}=\frac{1}{2} \left[\Gamma^2,\Gamma^1\right]=i\sigma_x.$$
Inserting $\hbar$ and $v_F$ we arrive at the following Hamiltonian
\begin{equation}
	\begin{split}
		-i \hbar \sigma_x \partial_t \psi = v_F \left[  \frac{1}{\sqrt{1+{{R}'}^2}} \sigma_z \left\{i p_z+\frac{\hbar R'}{2R}  \right\}  + \frac{\hbar}{R} \sigma_y \partial_\phi \right]\psi.
	\end{split}
\end{equation}
We can simplify by multiplying both sides by $-\sigma_x$ and get the following Hamiltonian 
\begin{equation}
	\begin{split}
		H &= v_F \left[ \frac{1}{\sqrt{1+{{R}'}^2}} \left\{i \hbar \partial_z + \frac{i \hbar {R}'}{2R}\right\} \sigma_y - \frac{i\hbar}{R} \sigma_z \partial_\phi\right].
	\end{split}
\end{equation}

\newpage
%\bibliographystyle{unsrt}
%\bibliography{lib2}


%

\end{document}